# The orientation of Julia Augusta Taurinorum (Torino)

**Amelia Carolina Sparavigna**
Department of Applied Science and Technology
Politecnico di Torino, C.so Duca degli Abruzzi 24, Torino, Italy

*It seems that the ancient Roman towns were oriented with the sunrise. Here I propose a discussion on the orientation of Torino, the Julia Augusta Taurinorum, which has the ancient Roman structure perfectly preserved. According to this ancient ritual, we can use the sunrise amplitude to determine the Turin's birthday. The use of the hour angle is also proposed, in this case the day of the foundation of Turin could be the winter solstice.*

Turin is located mainly on the left bank of the Po River, near the confluence with Dora River. A settlement in this area dated since the third century BC. It was known as the village of Taurasia of a Celto-Ligurian people, the Taurini. It seems that the name of this population is coming from a Celtic word meaning "mountain". According to some sources [1], Taurasia tried to impede the march of Hannibal when he was attacking Rome, coming from the Alps. For three days the town resisted, but was eventually destroyed by Hannibal. The origin of the modern city is in a castrum built by Julius Caesar during the Gallic Wars. In 27 BC, Torino became a Roman colony under the name of Julia Augusta Taurinorum.

The typical Roman street grid is clearly visible in the modern city, especially in that part known as the Quadrilatero Romano. In Ref. 2, it is told that the Roman town was a "centuriation", that is a land division, in the form of a rectangle of 770 m × 710 m, subdivided in 72 insulae (blocks). We can see in the Figure 1 that this structure is perfectly maintained. The "umbilicus", the center of the town, was at the crossing of the Decumanus Maximus and the Cardo Maximus, the two main streets. Via Garibaldi traces the exact path of the Decumanus, starting from the East gate, the Porta Praetoria now incorporated in Palazzo Madama, and ending at the West gate, the Porta Decumana. The Porta Palatina, on the north side of the town is still well preserved and is the origin of the Cardo Maximus. As we can see from a map dated 1572, drawn by Giovanni Caracha [3], the structure of the roman town persisted unaltered until the modern times [4], surrounded by the high walls built by Augustus.

The planning of the Roman town was performed by means of a centuriation [5], which was a method of land measurement and surveying. The centuriation is characterised by the regular grid traced using some surveyor's instruments. According to Ref.6, the foundation of a town followed a ritual, described by several Latin writers. The ritual comprised the observation of the flight of the birds and the outline of the perimeter by ploughing a furrow (as Romulus made for Rome). Ref.6 is telling that "the fundamental part of all the rituals of the aruspexes (priests in ancient Rome who practiced divination) was the individuation of the auguraculum, a sort of terrestrial image of the heavens (templum) in which the gods were "ordered" and "oriented" starting from north in the hourly direction. The individuation of the templum thus required astronomical orientation to the cardinal points." In fact, Ref.5 is telling that the surveyor first identified a central viewpoint, the "umbilicus agri" or "umbilicus soli". He then took up his position there and, looking towards the West, defined the place with the following names: ultra, the land in front of him, citra, the land behind him, dextera and sinistra, the land to his right and to his left.

Therefore, the roman town had generally the streets East-West oriented; the main one is the Decumanus. If the surveyor had as reference the place where the sun rose, this does not mean that the direction of the decumanus is perfectly the cardinal direction East-West (in agreement with the observation that "not all Roman centuriation displays consistent orientation [7]"). The decumanus can be inclined of a certain angle with respect the cardinal axis. In this case, measuring this angle of inclination we can give the day of the foundation.

In fact, Magli analyzed the orientations of the Roman towns in Italy to find any consistency with astronomical data [6]. He concluded that the orientation of these towns is not random. "It comprises two "families", one lying in the sector within ten degrees SE, the other near the winter solstice sunrise. Orientation of some towns to the sunrise in dates corresponding to important festivities of the Roman calendar, in particular Terminalia, looks also probable.  The existence of astronomical orientations confirms statements made by many Roman writers themselves, and raises the problem of the symbolic meaning of the castrum layout."

In Ref.6, Torino is not discussed. It is just given an angle of 30 (34) degrees. This is not the angle of Torino that we can measure on the satellite maps (see Fig.2). The actual angle the decumanus is forming with the cardinal East-West direction is 25.8 degrees (clockwise). Even considering that the sun has an apparent size of 1/2 degrees, the angle given in Ref.6 is wrong.

Now, let us consider the Turin Decumanus, Via Garibaldi, again. It is so straight, that we can see, within a few days of the summer solstice, the sun shining at the sunset on the building of Palazzo Madama, making the windowpanes glow like fires. During these days, we have also the celebrations and festival of Saint John the Baptist (June 24), who is the Patron Saint and protector of Torino. Considering this fact, and after reading [6], I searched for some references on the orientation of Torino. I found a book (in Italian, [8]), telling that the Decumanus is following the line of the ascendant sun, but nothing more. I prefer avoiding a discussion with astrological methods that I do not know; let us use the equations of wayfaring, that is, of those methods of orientation based on the observation of stars, sun and moon [9].

There are two quantities of orientation which are quite interesting functions of the solar declination. These quantities are the hour angle and the sunrise amplitude, the angle measured from East of the sunrise position on the horizon. In [9],   it is provided the solar declination as a function of the days after the spring equinox (for the angles in the horizontal and the equatorial coordinate systems, see [10]).

The hour angle (in degrees, evaluated with respect East-West direction) is given as a function of latitude φ and the solar declination δ:

$$\omega = \frac{360°}{2\pi} \arccos(-\tan\varphi \cdot \tan\delta) - 90° \qquad (1)$$

The declination (in radians) is:

$$\delta = \arcsin(0.4 \cdot \sin(2\pi n / 365)) \qquad (2)$$

In (2), n is representing the number of days after the spring equinox.
The sunrise amplitude, evaluate with respect the East-West direction) is:

$$Z = 90° - \frac{360°}{2\pi} \arccos(\sin\delta / \cos\varphi) \qquad (3)$$

For the umbilicus of Torino, φ  is equal to 45.07 degrees. Plotting (1) and (2) we have the Fig.3. From the plots we see that at Torino latitude, the angle of the decumanus (26° negative) corresponds to the sunrise amplitude on the days about 10 November or 30 January. If we imagine that Torino was founded with the Etruscan ritual, one to these two days is the foundation day.

In astronomic terms [10], the coordinate system used in this approach is the horizontal one. But, there is a problem: the horizon of Torino toward East is occupied by rather high hills. It seems therefore difficult that during the ritual, an observer could had seen the sunrise. Let us consider an alternative approach, based on the use of the equatorial coordinate system. From the plots in Fig.3, we see that it is quite interesting   the behavior of the hour angle, which on winter solstice is coincident with the angle of the decumanus. This means that the orientation of the decumanus could had been decided as a "local image of the sky", in a literal sense, where the angles of the main street of towns are equatorial angles. The maximum value of the hour angle is 25.95 degrees (positive for the summer solstice and negative for the winter solstice). At the winter solstice this hour angle turns out to be the angle of the decumanus of Turin. The birthday of town could be the winter solstice [11]. For Torino then, the name of the town was in honour of Julius Caesar, the orientation in honour of the Sun.

The same situation, a correspondence of the hour angle with the orientation of a structure, happens for the Great Temple of Amarna, where the orientation of the temple coincides with the hour angle on the winter solstice [12]. Other sites are under investigation to see whether orientation with equatorial coordinates is possible or not.

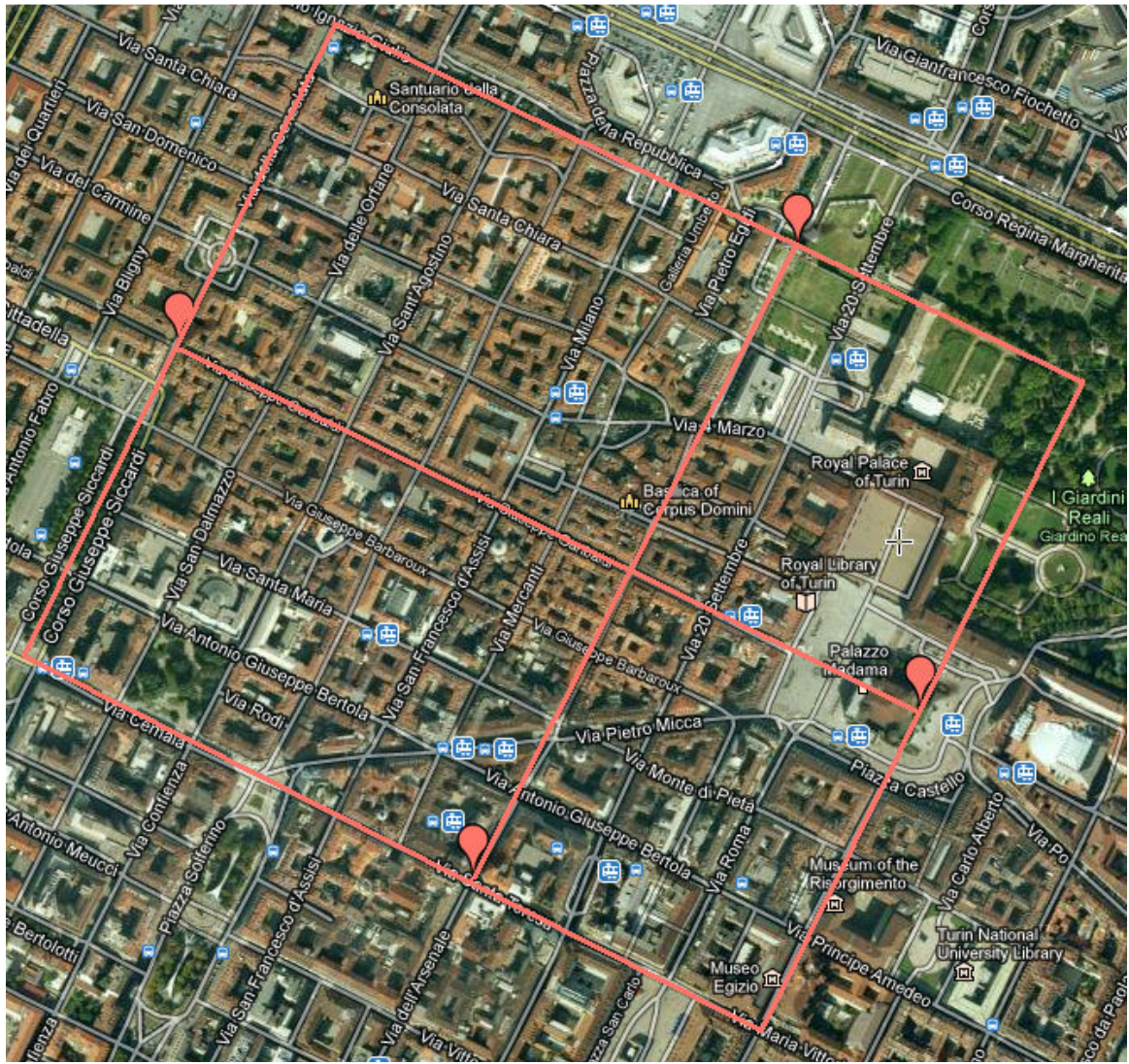

*Fig.1 This is the Roman Torino from Acme Mapper. The places of the four gates are marked (two of them are still existing). The Decumanus Maximus is the road inclined East-West line, coincident with the modern Via Garibaldi. Note the blocks coincident with the roman insulae. The "umbilicus", the centre of the town, is at the crossing of the main streets. The perimeter of the Roman town is going from Porta Palatina to Via della Consolata. Here the perimeter turned south on Via della Consolata and Corso Siccardi. On this side, there was the gate Porta Decumana, of which nothing remains. At the corner of Via Cernaia, the perimeter turns toward Porta Marmorea, completely dismantles. Today, on this path there are Via Cernaia, Santa Teresa, Maria Vittoria, Piazza San Carlo. At the corner of the Academy of Sciences, where we find the Egyptian Museum, the walls are running due North, crossing Piazza Castello, where there was the Porta Pretoria, then the area of the Royal Palace, returning to the Porte Palatine.*

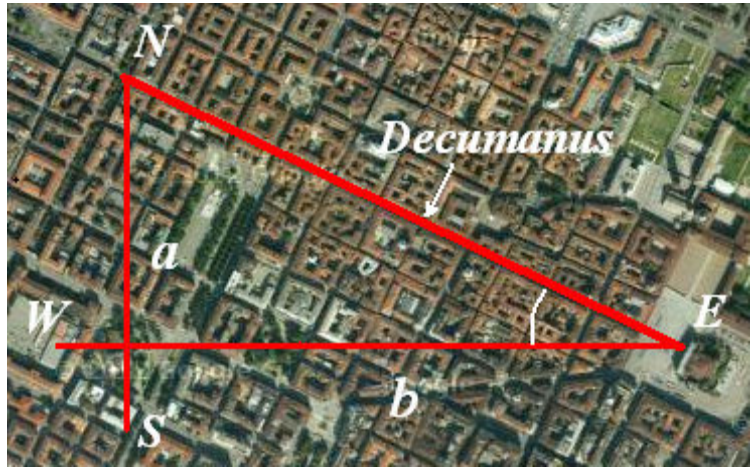

*Fig.2 The decumanus is the hypotenuse of the triangle. Measuring a and b we find the angle. From the figure we have an angle of 25.8 degrees.*

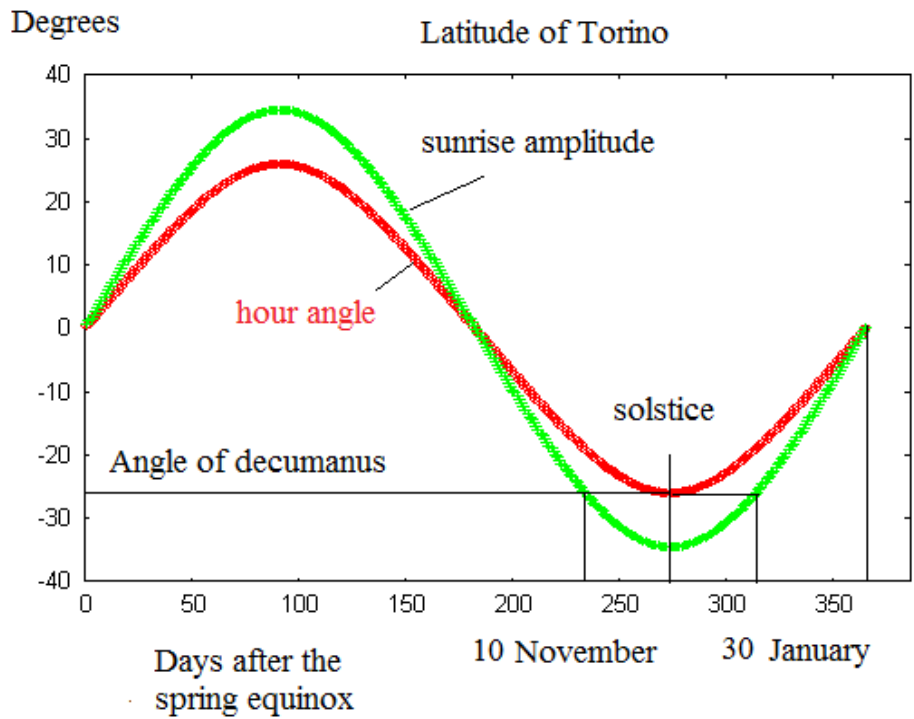

*Fig.3 Sunrise amplitude and hour angle at Torino as a function of the days from equinox. The angles are in degrees from East-West direction. The hour angle at the winter solstice is coincident with the orientation of the decumanus.*